\documentclass[10pt,aps,pra,twocolumn,superscriptaddress,nofootinbib,floatfix]{revtex4-2}
\usepackage[english]{babel}
\usepackage[utf8]{inputenc}
\usepackage[T1]{fontenc}
\usepackage{amsmath,amssymb,amsfonts}
\usepackage{graphicx}
\setkeys{Gin}{draft=false}
\usepackage{tikz}
\usetikzlibrary{arrows.meta,positioning,shapes.geometric}
\usepackage{booktabs}
\usepackage{tabularx}
\usepackage{adjustbox}
\usepackage{algpseudocode}

\usepackage[colorlinks=true,allcolors=blue]{hyperref}

\begin{document}

\title{Reverse Quantum Annealing for Ground-State Energy Calculation in Molecule}

\author{Pradyot Pritam Sahoo}
\email{pradyotpritam@gmail.com}
\affiliation{Centre for Quantum Engineering, Research and Education(CQuERE), TCG Crest, Kolkata 700091, India}
\affiliation{Academy of Scientific and Innovative Research (AcSIR), Ghaziabad 201002, India}

\date{\today}

\begin{abstract}
We present the results of our reverse quantum annealing computations of the ground-state energies of alkaline-earth monofluoride molecules. By partitioning the Dirac--Coulomb electronic Hamiltonian into sub-QUBOs and refining forward-annealing solutions through optimized reverse-annealing schedules, residual configuration errors are systematically minimized. The electronic ground-state wave functions and energies are obtained using the Quantum Annealing Eigensolver algorithm. This work provides a pathway for achieving sub-milliHartree chemical accuracy and advancing molecular electronic structure calculations in the quantum annealing paradigm.

\end{abstract}

\maketitle

\section{Introduction}

Accurate ground-state energies are central to molecular electronic-structure theory 
because they provide the energetic reference for bonding, stability, and molecular 
properties. Quantum annealing (QA) offers a hardware-based optimization paradigm in which 
a problem Hamiltonian is reached from an initial quantum Hamiltonian through a controlled 
annealing schedule\cite{Kadowaki1998}. For molecular electronic-structure problems, QA can be incorporated 
into a hybrid eigensolver in which the electronic Hamiltonian is mapped to a binary 
optimization problem and approximately optimized on a quantum annealer(D-Wave Quantum 
annealer).

A conventional forward-annealing (FA) calculation starts from a state dominated by 
quantum fluctuations and evolves toward the problem Hamiltonian. Practical devices, 
however, operate with finite annealing times and are affected by thermal fluctuations, 
embedding overhead, finite connectivity, chain-related effects, and stochastic sampling. 
Consequently, the sampled state need not coincide with the global minimum of the encoded 
problem, particularly as the size of the binary optimization problem increases.

Reverse quantum annealing (RQA) provides a complementary refinement mechanism. Instead of 
starting from the transverse-field-dominated initial state, RQA starts from a classical 
candidate configuration and temporarily moves toward a more quantum regime before 
returning to the problem Hamiltonian. The resulting local quantum-assisted search can, in 
favorable cases, move the system out of an excited-state basin while retaining useful 
information from the starting configuration.

Here we apply RQA to the Quantum Annealing Eigensolver (QAE) for relativistic molecular 
energy calculations of open-shell diatomic molecules. We study the alkaline-earth 
monofluorides BeF, MgF, CaF, SrF, and BaF in $(3e,8o)$ and $(7e,14o)$ active spaces. 
These open-shell molecules serve as benchmark systems in precision relativistic chemistry 
and tests of fundamental physics.
The QAE is broadly applicable to a wide range of 
physical optimization problems, 
including molecular vibrational spectra~\cite{Teplukhin2019},  energy calculations for 
molecular electronic states, relativistic fine-structure splitting in  atomic ions~\cite{vikrant2024}, computations of the permanent electric dipole moments~\cite{sahoo2026}.
 Moreover, these molecules are important for various applications, 
particularly for SrF~\cite{pra2014}, the first molecule to be laser-cooled~\cite{Shuman2010} and BaF, a promising candidate for precision measurements in fundamental physics~\cite{Aggarwal_2018}.

Twelve reverse-annealing schedules\cite{Barzegar2024} are examined, 
spanning pause-type and V-shaped trajectories across varying reversal depths and dwell 
times. We investigate: (i) how significantly RQA lowers the forward-annealing energy, 
(ii) which schedule parameters optimize quantum tunneling out of local traps, and (iii) 
how refinement scales as the active space expands.
Benchmarking against exact Full Configuration Interaction (FCI) diagonalization 
demonstrates that reverse annealing eliminates $82.3\%$ to $98.1\%$ of residual forward-
annealing errors, achieving sub-milliHartree chemical accuracy. The remainder of this 
paper is organized as follows: Section~\ref{sec:theory} details the relativistic 
Hamiltonian and QAE mapping; Section~\ref{sec:results} presents schedule designs and FCI 
benchmarks; and Section~\ref{sec:conclusions} concludes.

\section{Theory and Computational Details}
\label{sec:theory}

\subsection{Dirac--Coulomb Hamiltonian}

The relativistic electronic Hamiltonian is taken in the Dirac--Coulomb (DC) form,
\begin{equation}
\begin{aligned}
\hat{H}_{DC} = & \sum_i
\left[
c\,\boldsymbol{\alpha}\cdot\mathbf{p}_i
+\beta mc^2
-\sum_A\frac{Z_A}{|\mathbf r_i-\mathbf R_A|}
\right] \\
& +\sum_{i<j}\frac{1}{|\mathbf r_i-\mathbf r_j|},
\end{aligned}
\end{equation}
where $c$ is the speed of light, $\boldsymbol{\alpha}$ and $\beta$ are the Dirac matrices, $\mathbf p_i$ is the momentum of electron $i$, and $\mathbf r_i$ and $\mathbf R_A$ are electronic and nuclear coordinates.

In second quantization,
\begin{equation}
\hat H=
\sum_{pq}h_{pq}a_p^\dagger a_q+
\frac12\sum_{pqrs}g_{pqrs}
a_p^\dagger a_q^\dagger a_s a_r ,
\end{equation}
where $h_{pq}$ and $g_{pqrs}$ are one- and two-electron integrals in a relativistic spinor basis. The electronic eigenvalue problem is
\begin{equation}
\hat H|\Psi\rangle=E|\Psi\rangle .
\end{equation}
Using a finite determinant basis,
\begin{equation}
|\Psi\rangle=\sum_i c_i|\Phi_i\rangle ,
\end{equation}
where the determinants include the Dirac--Fock reference and selected particle--hole excitations~\cite{slatercondonrules}.

\subsection{Classical Reference Calculations}

Dirac--Fock (DF) and relativistic coupled-cluster singles and doubles (RCCSD) 
calculations~\cite{Dyall2007} were performed with DIRAC22~\cite{DIRAC22}. The DF 
calculation was performed at the self-consistent-field level, using the molecular double-group symmetry available in the calculations. Exact reference energies were generated by 
constructing the Full Configuration Interaction (FCI) Hamiltonian in the Slater--Condon 
determinant basis~\cite{slatercondonrules} and diagonalizing the resulting matrix via 
full exact diagonalization for all five molecular radicals across both $(3e,8o)$ and 
$(7e,14o)$ active spaces. These exact FCI eigenvalues serve as an absolute reference to 
evaluate the convergence of both forward-annealing and reverse-annealing states.

\subsection{Basis Sets and Numerical Details}

All relativistic calculations used uncontracted Gaussian-type orbitals with kinetic 
balance to maintain the appropriate coupling between the large and small components of 
the Dirac spinors~\cite{Dyall2007}. For Be, Mg, Ca and F, exponents were taken from 
correlation-consistent polarized valence basis sets obtained from the EMSL Basis Set 
Exchange~\cite{BSE}. For Sr and Ba, the dyall.c4v basis sets was employed, with 
additional tight primitives introduced where required~\cite{Dyall:2009,Dunning:1989}. The 
resulting relativistic one- and two-electron integrals define the molecular Hamiltonians 
used in the QAE calculations.

The molecular calculations were carried out for fixed active spaces of $(3e,8o)$ and 
$(7e,14o)$. The same electronic-structure and QAE construction was retained when 
comparing forward and reverse annealing so that the reported $\Delta E$ isolates the 
change associated with the annealing/refinement stage rather than a change in the 
underlying molecular Hamiltonian.

\subsection{Quantum Annealing Eigensolver}
QA involves preparing the ground-state of an initial Hamiltonian, $\hat{H}_I$, which is typically chosen to be the transverse field Hamiltonian, and then adiabatically transforming it into problem Hamiltonian, $\hat{H}_F$, generally chosen to be the Ising Hamiltonian whose ground-state encodes the solution to the problem. Where the system evolves smoothly from the ground state of $\hat{H}_I$ to that of $\hat{H}_F$ during the annealing time $T$. Initially, annealer starts with giving a simple delocalized ground state with spins. As time progresses, $\hat{H}_I$  decreases and $\hat{H}_F$ increases, and by $t = T$,  completing the annealing process.
While the adiabatic model described above is the theoretical foundation of QA, in practice, quantum annealing operates far from the adiabatic limit.
The QAE converts the eigenvalue problem into an optimization problem. Introducing a Lagrange multiplier $\lambda$ for normalization gives
\begin{equation}
\xi=\langle\Psi|\hat H|\Psi\rangle-\lambda\langle\Psi|\Psi\rangle .
\end{equation}
After expansion in the determinant basis,
\begin{equation}
\xi(\vec c,\lambda)=
\sum_{i,j=1}^{N}c_i c_j
\left(H_{ij}-\lambda\delta_{ij}\right),
\end{equation}
where
\begin{equation}
H_{ij}=\langle\Phi_i|\hat H|\Phi_j\rangle .
\end{equation}
The active spaces considered are
\begin{equation}
(N_o,M_e)=(8o,3e),\qquad (N_o,M_e)=(14o,7e),
\end{equation}
where $N_o$ is the number of active orbitals and $M_e$ is the number of active electrons.

\subsection{Binary Encoding and QUBO Construction}

Each coefficient $c_i\in[-1,1]$ is represented using $K$ binary variables $q_\alpha^i\in\{0,1\}$:
\begin{equation}
c_i=\sum_{\alpha=1}^{K-1}2^{\alpha-K}q_\alpha^i-q_K^i .
\end{equation}
Substitution into the QAE functional produces a QUBO,
\begin{equation}
\xi(\vec q,\lambda)=
\sum_{i,j=1}^{N}\sum_{\alpha,\beta=1}^{K}
Q_{i\alpha,j\beta}q_\alpha^i q_\beta^j .
\end{equation}
Thus the binary representation contains $NK$ variables for $N$ coefficients as explained in appendix~\ref{app:fixed_point_encoding}. The calculations use $K=10$ binary variables per coefficient and the D-Wave Ocean software environment~\cite{DwaveOceanSDK, DWaveOcean2022}.

\subsection{Sub-QUBO Decomposition and Forward Annealing}

Because finite hardware connectivity limits the size and density of directly embeddable problems, the full QUBO is partitioned into smaller sub-QUBOs~\cite{sahoo2026,vikrant2024}. Following the ordering strategy used in this workflow, coefficients are ordered according to the first order perturbation theory approximation,
$C_\alpha \sim H_{0\alpha}/(H_{00}-H_{\alpha\alpha})$ and their  amplitudes and grouped into computationally tractable blocks\cite{sahoo2026}. For this calculations, we selected energy ranges $[\lambda_{\text{min}}, 
\lambda_{\text{max}}]$ slightly below and above the DF
energy, corresponding to the active spaces $(8o,3e)$ and $(14o,7e)$ used in this study. 
This choice ensures
that the physically relevant region of the energy landscape is sampled while maintaining
numerical stability of the optimization. Each sub-QUBO is embedded on the Pegasus topology~\cite{Boothby2019,Pelofske_2024} and sampled on the quantum annealer. The lowest-energy sampled configuration is retained, with the hardware embedding/chain~\cite{Pelofske_2024} procedure following the implementation used in the present workflow~\cite{DWaveChainStrength:2022, DwaveOceanSDK} as explained in appendix~\ref{app:dwave_protocol}.

The sub-QUBO solutions are concatenated,
\begin{equation}
\vec q_{\rm raw}=
[\vec q_{\rm fwd}^{(1)},\vec q_{\rm fwd}^{(2)},\ldots,
\vec q_{\rm fwd}^{(M)}],
\end{equation}
to form the initial classical configuration for RQA. Because the fragments are optimized separately, $\vec q_{\rm raw}$ need not be the global minimum of the original QUBO.

\subsection{Reverse Quantum Annealing}

In RQA, the anneal parameter is taken from the problem-Hamiltonian endpoint to a reversal point $s'$ and then returned:
\begin{equation}
s:1\rightarrow s'\rightarrow1,\qquad 0<s'<1 .
\end{equation}
The reversal increases the relative importance of quantum fluctuations and enables exploration around the supplied state in our case forward annealing solution. The reversal depth is controlled by $s'$, while the dwell or excursion time controls the duration of the exploration~\cite{Mehta2025}.

\begin{figure*}[t]
\centering
\includegraphics[width=0.92\textwidth]{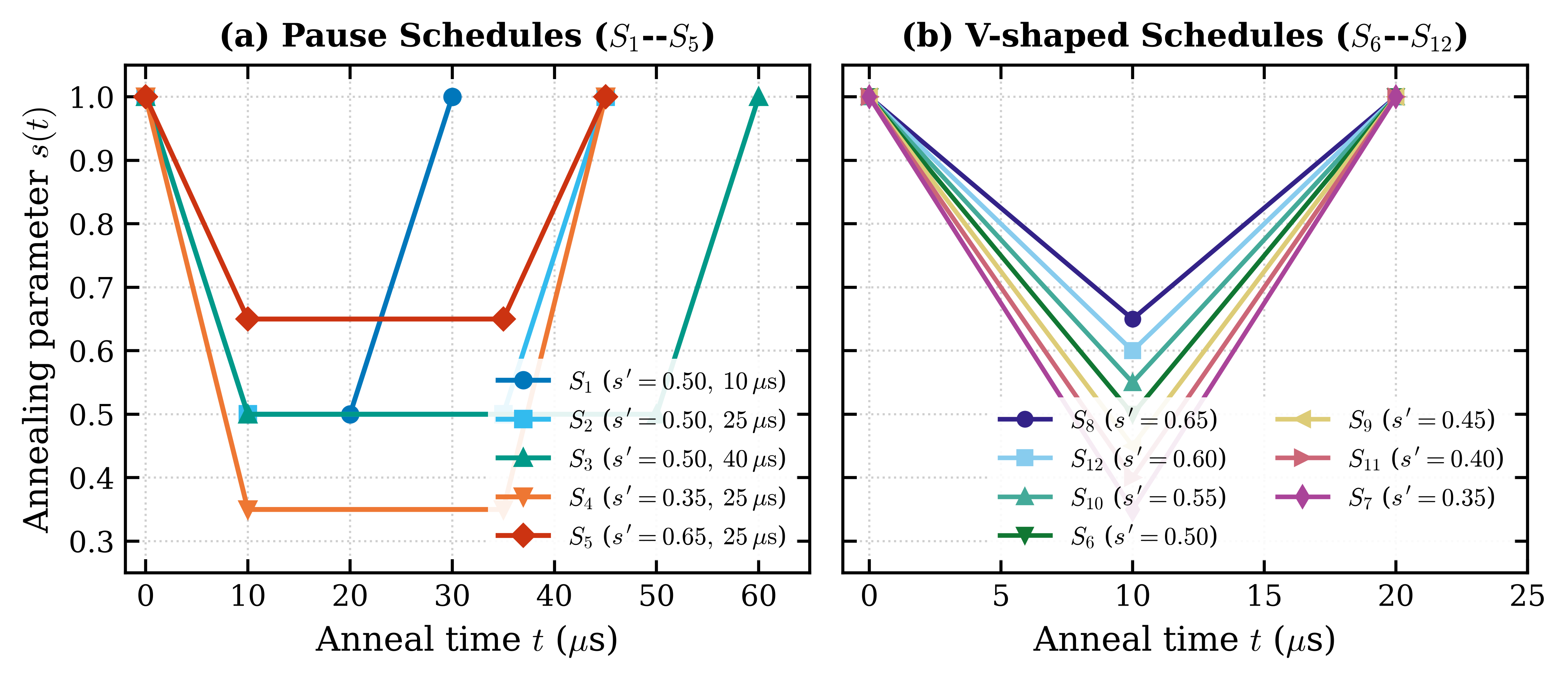}
\caption{Twelve reverse-annealing schedules evaluated in this work: (a) Pause schedules ($S_1$--$S_5$) with dwell intervals of $10$, $25$, and $40\,\mu\text{s}$ at intermediate reversal points $s'\in\{0.35, 0.50, 0.65\}$; (b) V-shaped schedules ($S_6$--$S_{12}$) with a fixed total anneal time of $20\,\mu\text{s}$ and reversal depths ranging from $s'=0.35$ to $s'=0.65$.}
\label{fig:schedules}
\end{figure*}

Twelve schedules, S1--S12, were used, following the reverse-annealing refinement strategy considered in recent work~\cite{manpreet2025}. S1--S5 are pause-type schedules, whereas S6--S12 are V-shaped schedules as shown in Figure~\ref{fig:schedules}. Their control points are given in Table~\ref{tab:schedules}. The schedule family spans reversal points from $s'=0.35$ to $s'=0.65$, together with different dwell times. This design permits the influence of reversal depth and schedule shape to be assessed without changing the molecular Hamiltonian or active space.

\begin{table}[t]
\centering
\caption{Twelve reverse-annealing schedules used in this work. Each schedule is defined by control points $[t,s(t)]$, with anneal time $t$ in $\mu\text{s}$.}
\label{tab:schedules}
\begin{ruledtabular}
\begin{tabular}{ll}
ID & Schedule points $[t,s(t)]$\\
\colrule
S1  & $[0,1],[10,0.50],[20,0.50],[30,1]$\\
S2  & $[0,1],[10,0.50],[35,0.50],[45,1]$\\
S3  & $[0,1],[10,0.50],[50,0.50],[60,1]$\\
S4  & $[0,1],[10,0.35],[35,0.35],[45,1]$\\
S5  & $[0,1],[10,0.65],[35,0.65],[45,1]$\\
S6  & $[0,1],[10,0.50],[20,1]$\\
S7  & $[0,1],[10,0.35],[20,1]$\\
S8  & $[0,1],[10,0.65],[20,1]$\\
S9  & $[0,1],[10,0.45],[20,1]$\\
S10 & $[0,1],[10,0.55],[20,1]$\\
S11 & $[0,1],[10,0.40],[20,1]$\\
S12 & $[0,1],[10,0.60],[20,1]$\\
\end{tabular}
\end{ruledtabular}
\end{table}

\subsection{Post-Processing}

The refined binary configuration obtained from RQA is decoded into real-valued coefficients. This step restores couplings that may be lost during sub-QUBO partitioning and yields the final variational energy.

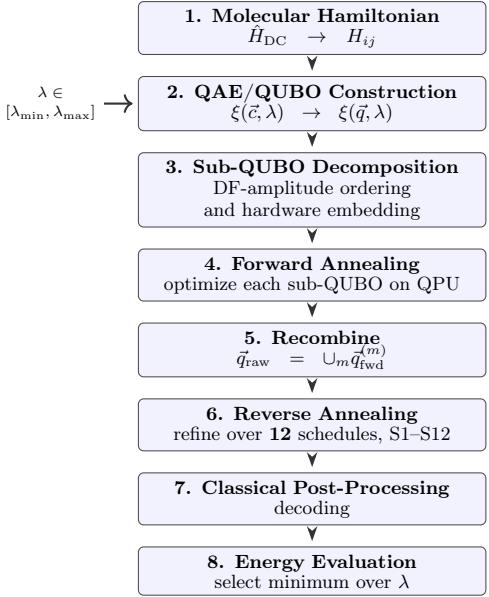
\begin{figure*}[t]
\centering
\begin{minipage}[t]{0.45\textwidth}
\vspace{0pt}
\centering
\begin{tikzpicture}[
    scale=0.80, every node/.style={transform shape},
    stage/.style={
        draw=black!70,
        rounded corners=2pt,
        fill=blue!5,
        align=center,
        inner sep=3.5pt,
        text width=5.5cm,
        minimum height=0.7cm,
        font=\small
    },
    myarrow/.style={
        -{Stealth[length=2.0mm,width=1.4mm]},
        thick,
        draw=black!80,
        shorten >=2pt,
        shorten <=2pt
    }
]
\node[stage] (ham) {\textbf{1. Molecular Hamiltonian}\\[-1pt]
$\hat H_{\rm DC}\rightarrow H_{ij}$};
\node[stage, below=0.35cm of ham] (qubo) {\textbf{2. QAE/QUBO Construction}\\[-1pt]
$\xi(\vec c,\lambda)\rightarrow\xi(\vec q,\lambda)$};
\node[stage, below=0.35cm of qubo] (subqubo) {\textbf{3. Sub-QUBO Decomposition}\\[-1pt]
DF-amplitude ordering and hardware embedding};
\node[stage, below=0.35cm of subqubo] (forward) {\textbf{4. Forward Annealing}\\[-1pt]
optimize each sub-QUBO on QPU};
\node[stage, below=0.35cm of forward] (combine) {\textbf{5. Recombine}\\[-1pt]
$\vec q_{\rm raw}=\cup_m\vec q^{(m)}_{\rm fwd}$};
\node[stage, below=0.35cm of combine] (reverse) {\textbf{6. Reverse Annealing}\\[-1pt]
refine over \textbf{12} schedules, S1--S12};
\node[stage, below=0.35cm of reverse] (post) {\textbf{7. Classical Post-Processing}\\[-1pt]
decoding };
\node[stage, below=0.35cm of post] (energy) {\textbf{8. Energy Evaluation}\\[-1pt]
select minimum over $\lambda$};

\draw[myarrow] (ham)--(qubo);
\draw[myarrow] (qubo)--(subqubo);
\draw[myarrow] (subqubo)--(forward);
\draw[myarrow] (forward)--(combine);
\draw[myarrow] (combine)--(reverse);
\draw[myarrow] (reverse)--(post);
\draw[myarrow] (post)--(energy);
\draw[myarrow, <-] (qubo.west) -- ++(-0.65,0)
    node[left, font=\scriptsize, align=center, inner sep=1pt] {$\lambda \in$\\[1pt]$[\lambda_{\min},\lambda_{\max}]$};
\end{tikzpicture}
\end{minipage}\hfill
\begin{minipage}[t]{0.52\textwidth}
\vspace{0pt}
\hrule\vspace{4pt}
\noindent\textbf{Algorithm 1:} Reverse-QAE execution protocol\label{alg:reverse_qae}
\vspace{3pt}\hrule\vspace{4pt}
\begin{algorithmic}[1]
\Require Relativistic Hamiltonian $H$, binary precision $K$, scan interval $\lambda\in[\lambda_{\min},\lambda_{\max}]$, schedules $\{S_i\}_{i=1}^{12}$
\Ensure Optimal ground-state eigenvector $\vec c_{\rm final}$ and energy $E_{\rm min}$

\State Formulate continuous QAE functional $\xi(\vec c,\lambda)$
\State Map $\vec c\in[-1,1]^N$ to $NK$ binary spins $\vec q\in\{0,1\}^{NK}$
\State Partition QUBO into sub-QUBO blocks $\{\xi_m\}_{m=1}^M$
\State Compute minor embeddings onto D-Wave Pegasus graph
\For{$\lambda \in [\lambda_{\min},\lambda_{\max}]$}
    \For{$m=1$ to $M$}
        \State Perform forward quantum annealing on sub-QUBO $\xi_m$
        \State Retain lowest-energy sample $\vec q_{\rm fwd}^{(m)}$
    \EndFor
    \State Form initial classical configuration $\vec q_{\rm raw} \gets \bigcup_{m=1}^M \vec q_{\rm fwd}^{(m)}$
    \For{$i=1$ to $12$}
        \State $\vec q_{\rm RA}^{(i)} \gets \mathrm{ReverseAnneal}(\vec q_{\rm raw}, S_i)$
        \State Decode configuration and evaluate energy
    \EndFor
    \State $\vec q_{\rm best}(\lambda) \gets \arg\min_i \xi(\vec q_{\rm RA}^{(i)},\lambda)$
    \State Decode binary state $\vec q_{\rm best}(\lambda) \rightarrow \vec c_{\rm best}(\lambda)$
    \State Evaluate variational energy \\ $E(\lambda) = \langle \Psi(\vec c_{\rm final}) | \hat{H}_{\rm DC} | \Psi(\vec c_{\rm final}) \rangle$
\EndFor
\State \Return $\vec c_{\rm final} \gets \arg\min_\lambda E(\lambda)$ and $E_{\rm min} \gets \min_\lambda E(\lambda)$
\end{algorithmic}
\vspace{4pt}\hrule
\end{minipage}
\caption{Computational architecture of the reverse-QAE algorithm. Left: Workflow schematic detailing the transformation from the relativistic Dirac--Coulomb Hamiltonian to the partitioned QUBO, forward annealing, multi-schedule reverse annealing, and continuous post-processing. Right: Algorithmic procedure executed across the normalization multiplier $\lambda$ to obtain the variational ground-state energy.}
\label{fig:rqae_workflow}
\end{figure*}

\section{Results and Discussion}
\label{sec:results}

\subsection{Definition of the RQA Improvement}

To quantify refinement relative to forward annealing, we use the signed relative energy change
\begin{equation}
\Delta E=
\frac{E_{\rm RA}-E_{\rm FA}}{|E_{\rm FA}|}\times100 .
\label{eq:delta}
\end{equation}
With this convention, $\Delta E<0$ denotes an energy lowering by RQA, 
$\Delta E=0$ denotes no change, and $\Delta E>0$ denotes a higher RQA 
energy. The signed definition is preferable to an absolute difference 
because it distinguishes genuine improvements from forward annealing.

We analyzed all molecule schedule observations for each active space: 
five molecules across twelve schedules as shown in Figures ~ \ref{fig:supp_energy_3e8o}, \ref{fig:supp_energy_7e14o}. All schedule-level averages 
below are averages over the five molecules, while molecule-level 
averages are averages over the twelve schedules as explained in appendix~\ref{app:data-analysis}.

\subsection{Schedule Dependence}

The schedule dependence is pronounced in both active spaces. In 
$(3e,8o)$, S11 gives the largest mean energy lowering, $-0.690\%$, 
followed by S9 ($-0.617\%$), S3 ($-0.597\%$), and S1 ($-0.548\%$). The 
weakest nonzero improvement is obtained with S5 ($-0.039\%$), while S8 
reproduces the forward-annealing energy in our calculation.

In $(7e,14o)$, S11 again performs best, with a mean change of 
$-0.151\%$, followed by S9 ($-0.129\%$), S4 ($-0.099\%$), S7 
($-0.091\%$), and S2 ($-0.090\%$). S5, S8, and S12 show zero mean 
change in our calculation. Thus, the ranking is not identical between 
active spaces, but the most favorable region is consistently 
associated with intermediate-to-deep reversals rather than the 
shallowest schedules.

The strongest schedules have reversal points around $s'=0.35$--$0.45$. S11, with $s'=0.40$, is the best schedule in both active spaces, while S9, with $s'=0.45$, is second. This is consistent with a Goldilocks-type regime in which the transverse-field contribution is sufficiently strong to explore nearby configurations but the initial FA information is not completely erased.

\begin{table}[htbp]
\centering
\caption{Mean signed relative energy change, $\Delta E$, for each schedule. Values are averaged over the five molecules. Negative values indicate lowering relative to forward annealing.}
\label{tab:scheduler_results}
\begin{ruledtabular}
\begin{tabular}{lrr}
Schedule & $(3e,8o)$ & $(7e,14o)$\\
\colrule
S11 & $-0.690$ & $-0.151$\\
S9  & $-0.617$ & $-0.129$\\
S3  & $-0.597$ & $-0.076$\\
S1  & $-0.548$ & $-0.065$\\
S4  & $-0.381$ & $-0.099$\\
S2  & $-0.446$ & $-0.090$\\
S12 & $-0.416$ & $0.000$\\
S6  & $-0.407$ & $-0.066$\\
S7  & $-0.386$ & $-0.091$\\
S10 & $-0.380$ & $-0.087$\\
S5  & $-0.039$ & $0.000$\\
S8  & $0.000$ & $0.000$\\
\end{tabular}
\end{ruledtabular}
\end{table}

\begin{figure*}[t]
\centering
\includegraphics[width=0.92\textwidth]{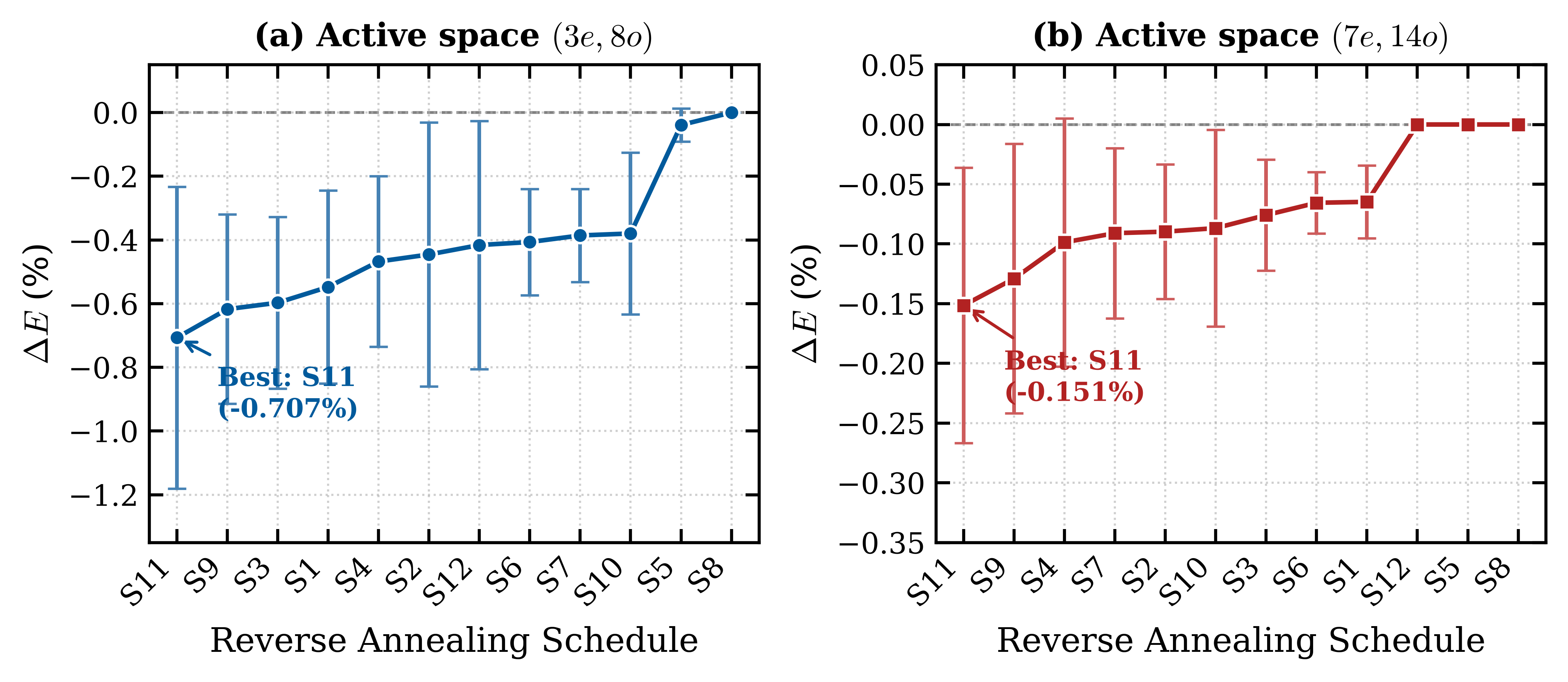}
\caption{Schedule-dependent RQA performance for (a) $(3e,8o)$ and (b) 
$(7e,14o)$ active spaces. Schedules are ordered monotonically from 
greatest average energy lowering to zero. Data points denote the mean 
signed relative energy change $\Delta E = (E_{\rm RA}-E_{\rm 
FA})/|E_{\rm FA}|\times 100$, and error bars denote the standard 
deviation ($\pm 1\sigma$) across the five molecular radicals. Schedule 
S11 ($s'=0.40$) achieves the highest mean energy lowering in both 
active spaces.}
\label{fig:sched_perf}
\end{figure*}

\subsection{Active-Space Dependence}

A major trend is the reduction in the recoverable RQA improvement when 
the active space is enlarged. Averaged over all molecule--schedule 
calculations, the signed relative changes are $-0.409\%$ for $(3e,8o)$ 
and $-0.071\%$ for $(7e,14o)$. At the best schedule, S11, the 
corresponding mean changes are $-0.690\%$ and $-0.151\%$.

The best-schedule improvement therefore decreases by a factor of 
approximately $4.5$ on moving from $(3e,8o)$ to $(7e,14o)$ which  
indicates that the particular QUBO encoding, sub-QUBO decomposition, 
embedding, and schedule set used here provide less recoverable 
improvement in the larger active space. The larger binary problem also 
introduces greater opportunities for embedding and stochastic errors\cite{Pearson2019} 
to influence the observed solution.

\subsection{Molecule Dependence}

The RQA response varies substantially among the five molecules. In 
$(3e,8o)$, MgF has the largest molecule-averaged lowering, $-0.568\%$, 
followed by BaF ($-0.475\%$), CaF ($-0.384\%$), SrF ($-0.356\%$), and 
BeF ($-0.262\%$). In $(7e,14o)$, the corresponding ordering is MgF 
($-0.138\%$), BeF ($-0.098\%$), CaF ($-0.061\%$), BaF ($-0.036\%$), 
and SrF ($-0.022\%$).


\begin{table}[htbp]
\centering
\caption{Molecule-averaged signed relative energy change. Values are averaged over the twelve schedules.}
\label{tab:molecule_results}
\begin{ruledtabular}
\begin{tabular}{lrr}
Molecule & $(3e,8o)$ & $(7e,14o)$\\
\colrule
BaF & $-0.475$ & $-0.036$\\
BeF & $-0.262$ & $-0.098$\\
CaF & $-0.384$ & $-0.061$\\
MgF & $-0.568$ & $-0.138$\\
SrF & $-0.356$ & $-0.022$\\
\end{tabular}
\end{ruledtabular}
\end{table}

\begin{figure*}[t]
\centering
\includegraphics[width=0.92\textwidth]{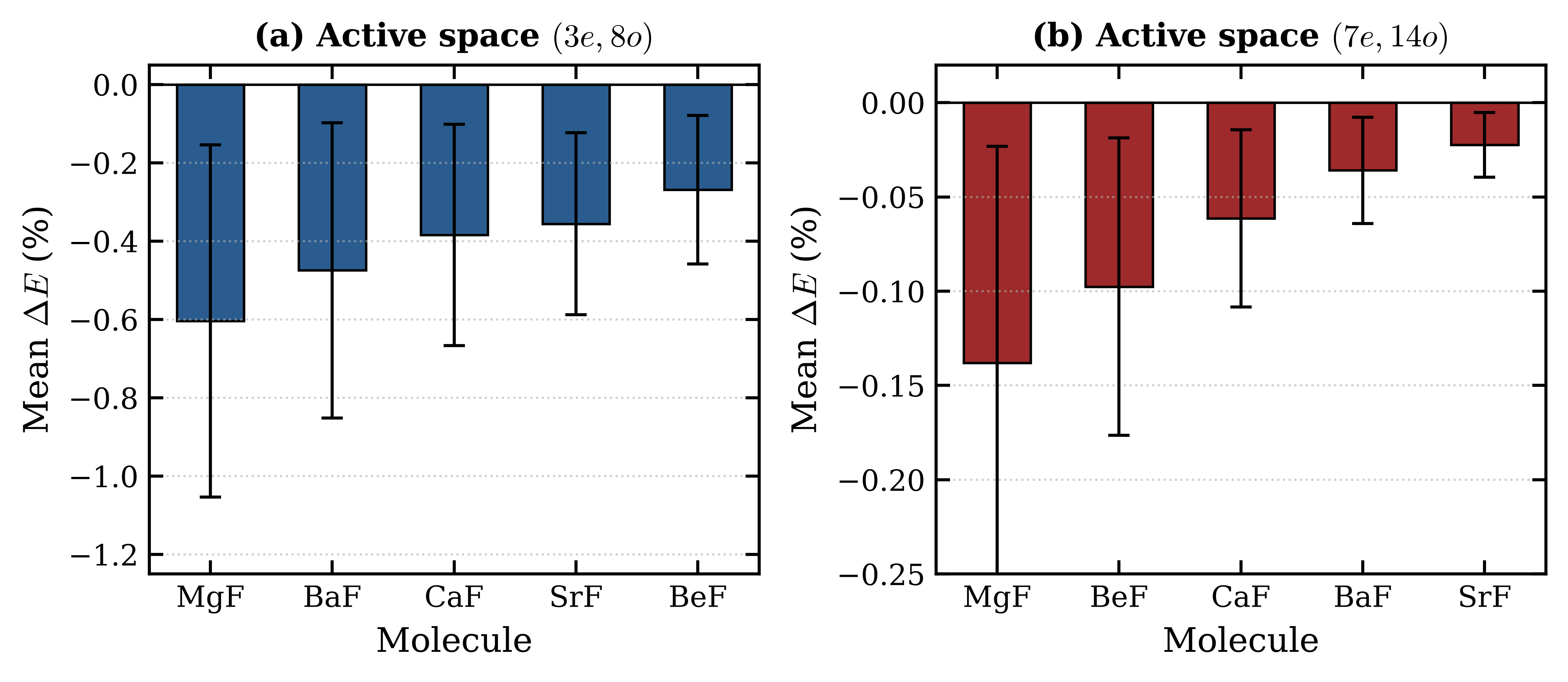}
\caption{Molecule-averaged signed RQA energy change for (a) $(3e,8o)$ 
and (b) $(7e,14o)$ active spaces. Negative values denote variational 
energy lowering relative to the forward-annealing baseline. Error bars 
indicate the standard deviation across the twelve reverse-annealing 
schedules.}
\label{fig:mol_imp}
\end{figure*}

\subsection{Best Schedules at the Molecule Level}

The best schedule is molecule dependent. In $(3e,8o)$, the lowest 
observed relative changes are obtained with S1 for BaF ($-1.054\%$) 
and BeF ($-0.554\%$), S4 for CaF ($-0.759\%$), and S11 for MgF 
($-1.292\%$) and SrF ($-0.712\%$). In $(7e,14o)$, S2 is best for BaF 
($-0.085\%$), S9 for BeF ($-0.225\%$), and S11 for CaF ($-0.139\%$), 
MgF ($-0.335\%$), and SrF ($-0.051\%$).

This molecule-level variation is important for practical schedule 
selection. It suggests that a small schedule ensemble may be more 
effective than assuming that one universal schedule is optimal for every 
molecular QUBO.

\subsection{Variability and Robustness}

RQA improvement is accompanied by stochastic variability. For S11, the 
standard deviation across molecules is approximately $0.505\%$ in 
$(3e,8o)$ and $0.115\%$ in $(7e,14o)$. The deeper schedules can 
therefore produce both larger mean improvements and broader molecule-
to-molecule responses.

Across all observations, RQA lowers the forward-annealing energy in 50 
cases in $(3e,8o)$, leaves it unchanged in 8 cases, and produces a 
higher energy in 2 cases. In $(7e,14o)$, 45 cases are improved, 15 are 
unchanged, and none are higher. The two higher-energy observations in 
$(3e,8o)$ are BeF/S11 ($+0.043\%$) and MgF/S4 ($+0.218\%$). All the 
RQA refined energy are given in supplementary material~
[\ref{fig:supp_energy_3e8o},\ref{fig:supp_energy_7e14o}].

\subsection{Quantitative Benchmark against Exact Full Configuration Interaction (FCI)}

A central problem for this  heuristic eigensolver is whether the energy 
lowering achieved by reverse quantum annealing represents true 
convergence toward the electronic ground state. To address this 
question rigorously, we benchmark the forward-annealing and reverse-
annealing variational energies against exact Full Configuration 
Interaction (FCI) diagonalization performed in the identical 
relativistic spinor determinant basis.

Table~\ref{tab:fci_benchmark} presents the complete FCI benchmark across both active spaces and all 
five molecular radicals. For each system, we compare the lowest forward-annealing energy $E_{\rm FA}$, 
the best reverse-annealing energy $E_{\rm RA}^{\rm best}$, the corresponding optimal schedule, the 
residual energy errors relative to FCI ($|E - E_{\rm FCI}|$ in milliHartrees), and the percentage 
reduction in residual error achieved by RQA:
\begin{equation}
\eta = \frac{|E_{\rm FA} - E_{\rm FCI}| - |E_{\rm RA}^{\rm best} - E_{\rm FCI}|}{|E_{\rm FA} - E_{\rm FCI}|} \times 100\%.
\end{equation}

\begin{table*}[t]
\centering
\caption{Benchmark against exact Full Configuration Interaction (FCI) calculations for alkaline-earth monofluorides in $(3e,8o)$ and $(7e,14o)$ active spaces. Energies are in Hartrees ($\text{E}_{\rm h}$); $\delta_{\rm FA} \equiv |E_{\rm FA}-E_{\rm FCI}|$ and $\delta_{\rm RA} \equiv |E_{\rm RA}^{\rm best}-E_{\rm FCI}|$ denote residual errors relative to exact FCI in milliHartrees ($\text{mHa}$); $\Delta E$ is the signed change relative to FA (\%); and $\eta$ is the percentage reduction in residual error achieved by RQA.}
\label{tab:fci_benchmark}
\begin{ruledtabular}
\setlength{\tabcolsep}{4.5pt}
\begin{tabular}{lcccccccc}
Molecule & $E_{\rm FCI}$ (Ha) & $E_{\rm FA}$ (Ha) & $E_{\rm RA}^{\rm best}$ (Ha) & Best Sched. & $\Delta E$ (\%) & $\delta_{\rm FA}$ (mHa) & $\delta_{\rm RA}$ (mHa) & $\eta$ (\%)\\
\colrule
\multicolumn{9}{c}{\textit{Active space $(3e,8o)$}}\\
BeF & $-2.879126$ & $-2.862953$ & $-2.878821$ & S1  & $-0.554$ & $16.173$ & $\mathbf{0.305}$ & $98.12\%$\\
MgF & $-2.548084$ & $-2.510064$ & $-2.542488$ & S11 & $-1.292$ & $38.021$ & $5.596$ & $85.28\%$\\
CaF & $-2.372715$ & $-2.351043$ & $-2.368891$ & S4  & $-0.759$ & $21.672$ & $3.825$ & $82.35\%$\\
SrF & $-2.318931$ & $-2.299062$ & $-2.315439$ & S11 & $-0.712$ & $19.869$ & $3.492$ & $82.42\%$\\
BaF & $-2.221598$ & $-2.194220$ & $-2.217353$ & S1  & $-1.054$ & $27.378$ & $4.246$ & $84.49\%$\\
\colrule
\multicolumn{9}{c}{\textit{Active space $(7e,14o)$}}\\
BeF & $-17.081188$ & $-17.035968$ & $-17.074222$ & S9  & $-0.225$ & $45.220$ & $6.966$ & $84.59\%$\\
MgF & $-16.011878$ & $-15.952295$ & $-16.005813$ & S11 & $-0.335$ & $59.583$ & $6.065$ & $89.82\%$\\
CaF & $-15.512783$ & $-15.488854$ & $-15.510336$ & S11 & $-0.139$ & $23.929$ & $2.447$ & $89.78\%$\\
SrF & $-15.254824$ & $-15.246472$ & $-15.254229$ & S11 & $-0.051$ & $8.352$  & $\mathbf{0.596}$ & $92.87\%$\\
BaF & $-14.740778$ & $-14.726527$ & $-14.739000$ & S2  & $-0.085$ & $14.251$ & $1.778$ & $87.53\%$\\
\end{tabular}
\end{ruledtabular}
\end{table*}

\begin{figure*}[t]
\centering
\includegraphics[width=0.92\textwidth]{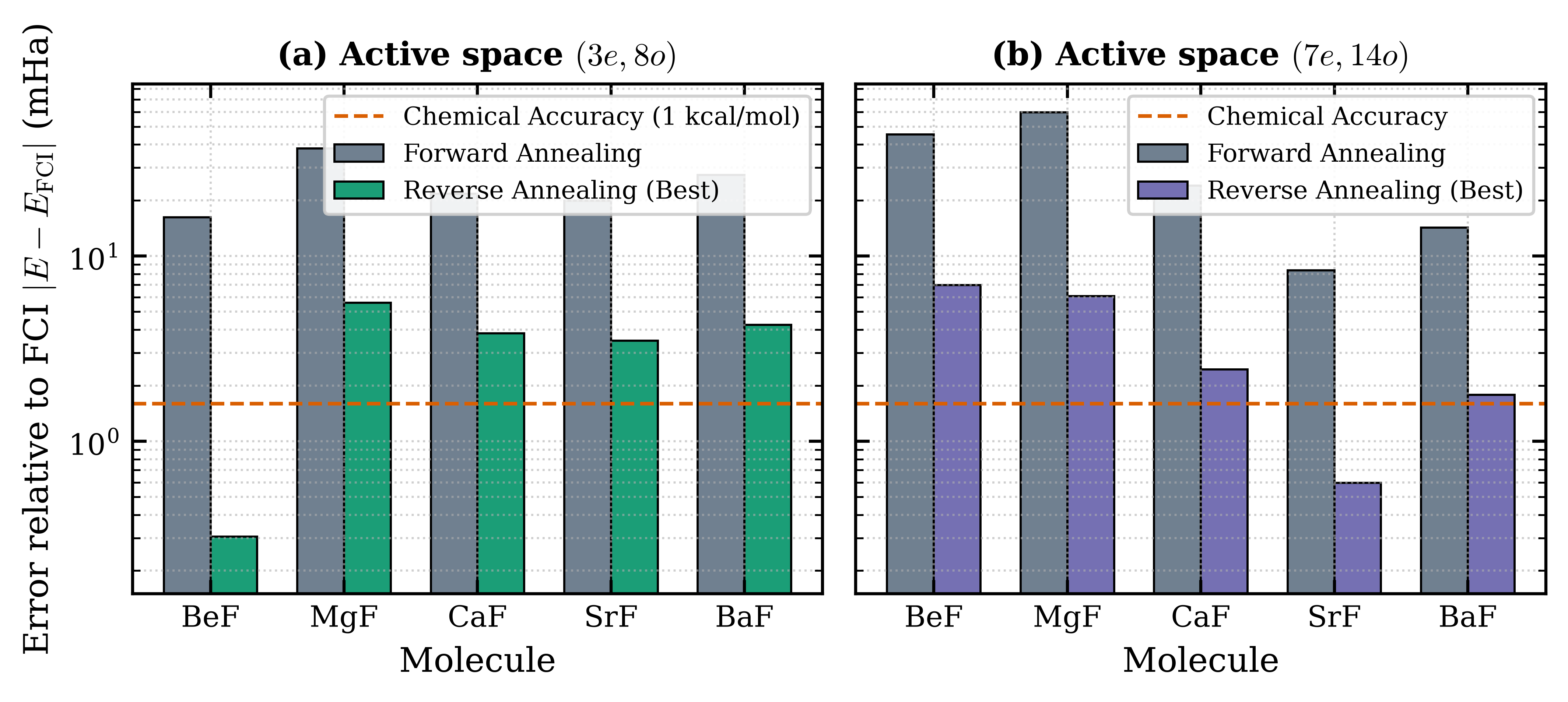}
\caption{Benchmark of residual ground-state energy errors $|E - E_{\rm FCI}|$ (in milliHartrees, logarithmic scale) relative to exact Full Configuration Interaction (FCI) for (a) $(3e,8o)$ and (b) $(7e,14o)$ active spaces. Forward annealing (FA, slate bars) leaves substantial residual configuration errors, whereas optimal reverse annealing (RA, colored bars) dramatically reduces the error by $82.3\%$ to $98.1\%$, reaching sub-milliHartree chemical accuracy ($<1.594\,\text{mHa} = 1.0\,\text{kcal/mol}$, dashed orange line) for BeF and SrF.}
\label{fig:fci_benchmark}
\end{figure*}

The benchmark demonstrates conclusively that RQA drives the solution systematically toward the exact ground state across all ten investigated systems:
\begin{enumerate}
\item \textit{Substantial error elimination:} RQA slashes the residual ground-state error by $82.3\%$ to $98.1\%$ across the five molecules. In the $(3e,8o)$ space, the residual error is reduced from $16.173$~mHa to $0.305$~mHa for BeF (a $98.12\%$ error reduction) and from $27.378$~mHa to $4.246$~mHa for BaF ($84.49\%$). In the $(7e,14o)$ space, the residual error is reduced from $45.220$~mHa to $6.966$~mHa for BeF ($84.59\%$), from $59.583$~mHa to $6.065$~mHa for MgF ($89.82\%$), and from $8.352$~mHa to $0.596$~mHa for SrF ($92.87\%$).
\item \textit{Attainment of chemical accuracy:} The standard threshold for chemical accuracy in quantum chemistry is $1.0~\text{kcal/mol} \approx 1.594~\text{mHa}$. As illustrated in Fig.~\ref{fig:fci_benchmark}, forward annealing leaves every molecule with errors well above the chemical accuracy line (ranging from $8.35$~mHa up to $59.58$~mHa). In contrast, RQA successfully brings BeF in $(3e,8o)$ ($0.305$~mHa) and SrF in $(7e,14o)$ ($0.596$~mHa) deep into the sub-milliHartree chemical accuracy regime. For BaF in $(7e,14o)$ ($1.778$~mHa), RQA brings the error to the threshold of chemical accuracy.
\item \textit{Physical interpretation:} The fact that RQA consistently lowers energy toward the exact FCI eigenvalue demonstrates that the transverse-field fluctuation introduced during the RQA cycle successfully unlocks trapped configuration basins and steers the system toward the true global minimum of the relativistic electronic Hamiltonian.
\end{enumerate}

In summary, rigorous benchmarking against exact FCI establishes that the combination of QAE sub-QUBO partitioning with reverse quantum annealing achieves high-precision relativistic quantum eigensolutions. By exploiting intermediate transverse-field fluctuations ($s'\approx 0.35$--$0.45$), RQA avoids the convergence bottlenecks  while overcoming the embedding and chain overhead of direct forward annealing on near-term hardware.

\section{Conclusions}
\label{sec:conclusions}

We investigated reverse quantum annealing as a refinement layer for QAE-based relativistic molecular energy calculations. Five alkaline-earth monofluorides (BeF, MgF, CaF, SrF, and BaF) were considered in $(3e,8o)$ and $(7e,14o)$ active spaces, and twelve reverse-annealing schedules were evaluated.

The results demonstrate that RQA performance is strongly schedule dependent, with S11 ($s'=0.40$) emerging as the most effective schedule on average in both active spaces, lowering the forward-annealing energy by $0.690\%$ and $0.151\%$, respectively. The optimal operational window corresponds to intermediate reversal depths ($s'\approx 0.35$--$0.45$). Benchmarking against exact Full Configuration Interaction (FCI) diagonalization further shows that RQA reduces residual forward-annealing errors by $82.3\%$ to $98.1\%$ across all five molecules. Sub-milliHartree chemical accuracy is achieved for BeF in $(3e,8o)$, with a residual error of $0.305$~mHa, and for SrF in $(7e,14o)$, with a residual error of $0.596$~mHa.

Upon enlarging the active space to $(7e,14o)$, the average recoverable relative improvement decreases from $0.409\%$ to $0.071\%$. Nevertheless, the absolute energy recovery remains substantial, eliminating approximately $85$--$93\%$ of the residual error relative to FCI. The response also depends on the molecular system, with MgF exhibiting the largest molecule-averaged energy lowering in both active spaces.

Overall schedule engineering is established by all these findings and reverse quantum annealing as effective refinement strategies for quantum-annealing-based relativistic electronic-structure calculations, offering a applied pathway toward the chemical accuracy on near-term quantum annealers.

\begin{acknowledgments}
The author gratefully acknowledges B. P. Das and Srinivasa Prasannaa V for valuable discussions and CQuERE, TCG CREST for providing access to D-Wave quantum annealing resources used in this work.
\vspace{5pt}
\end{acknowledgments}

\bibliographystyle{apsrev4-2}
\bibliography{ref}


\appendix

\section{Supplementary Material: Molecular Ground-State Energy Profiles and Schedule Benchmarks}
\subsection{Data-analysis definitions}
\label{app:data-analysis}
For each molecule and schedule, the supplied data contain a forward-annealing energy $E_{\rm FA}$ and a reverse-annealing energy $E_{\rm RA}$. The signed relative change used in this paper is
\begin{equation}
\Delta E =
\frac{E_{\rm RA}-E_{\rm FA}}{|E_{\rm FA}|}\times100.
\end{equation}
Schedule averages are arithmetic means over the five molecular systems, while molecule averages are arithmetic means over the twelve schedules. Standard deviations are computed over the corresponding five or twelve observations, respectively.

\label{app:supplementary_profiles}

This supplementary section presents the complete molecule-resolved energy evaluations across all twelve reverse-annealing schedules ($S_1$--$S_{12}$), comparing forward-annealing (FA), reverse-annealing (RA), and exact Full Configuration Interaction (FCI) ground-state reference values.

\subsection{Molecule-Resolved Energy Profiles in the $(3e,8o)$ Active Space}

\begin{figure*}[t]
\centering
\begin{minipage}{0.31\textwidth}
  \centering
  \includegraphics[width=\linewidth]{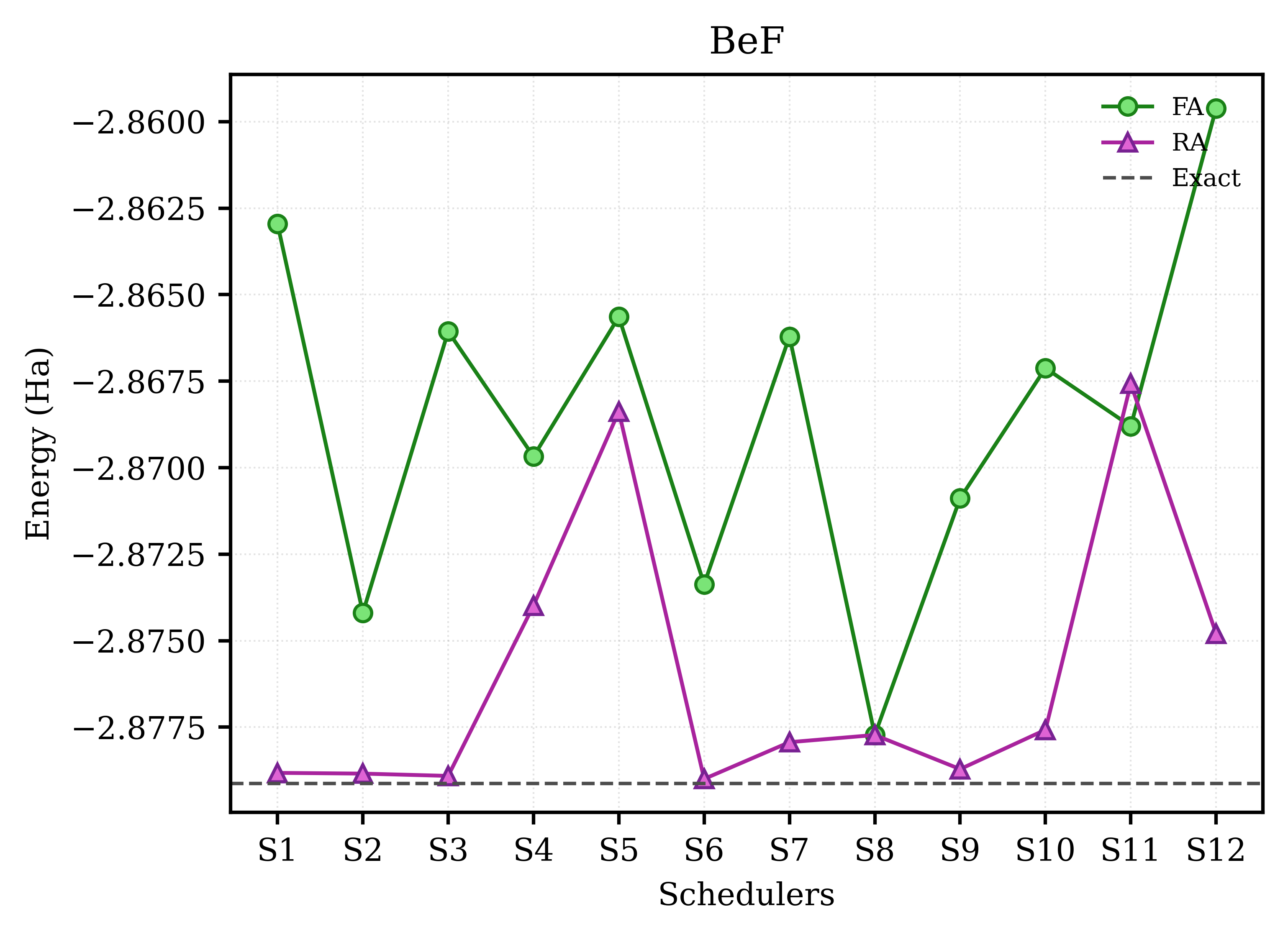}\\
  \footnotesize (a) BeF $(3e,8o)$
\end{minipage}\hfill
\begin{minipage}{0.31\textwidth}
  \centering
  \includegraphics[width=\linewidth]{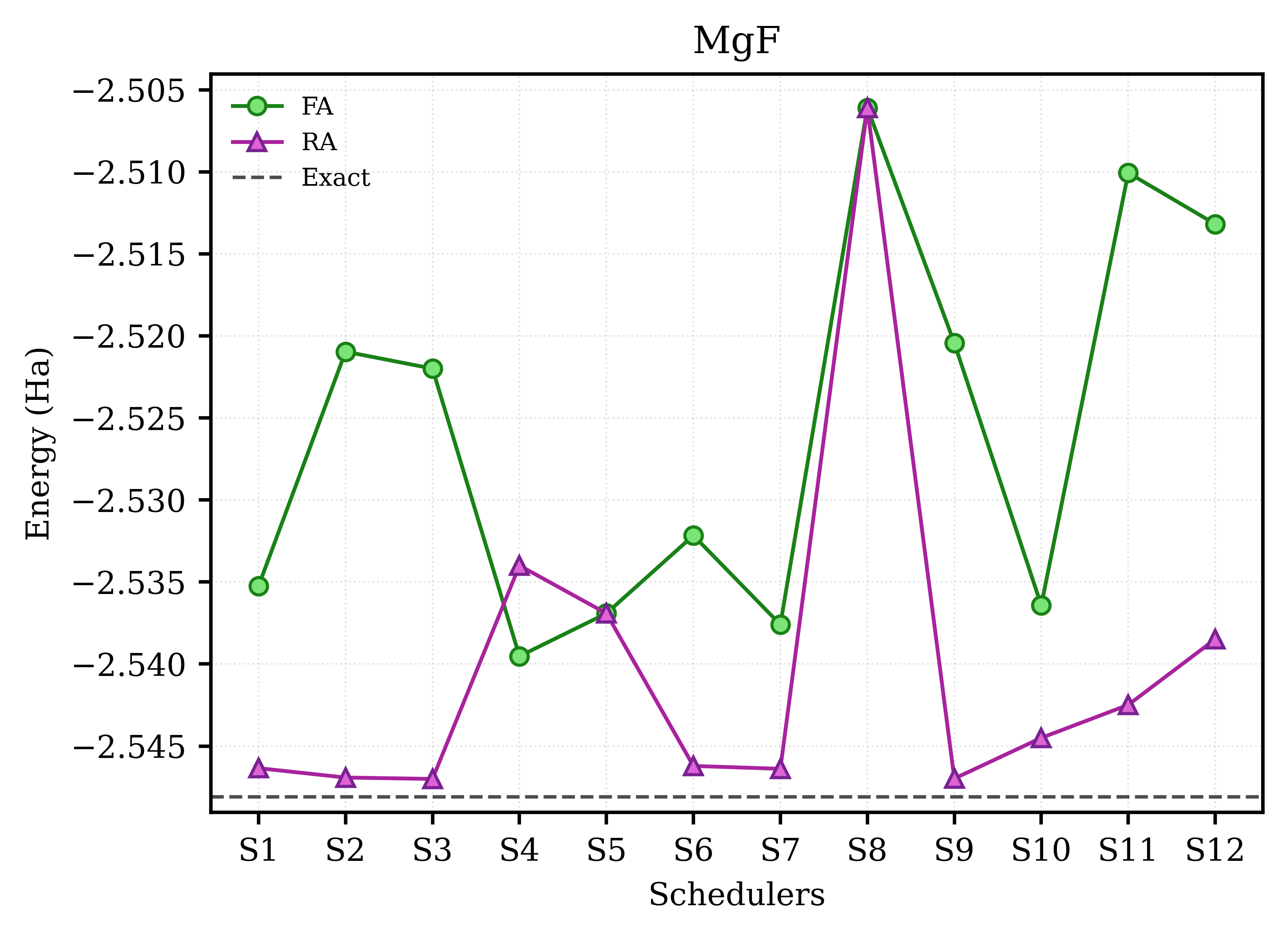}\\
  \footnotesize (b) MgF $(3e,8o)$
\end{minipage}\hfill
\begin{minipage}{0.31\textwidth}
  \centering
  \includegraphics[width=\linewidth]{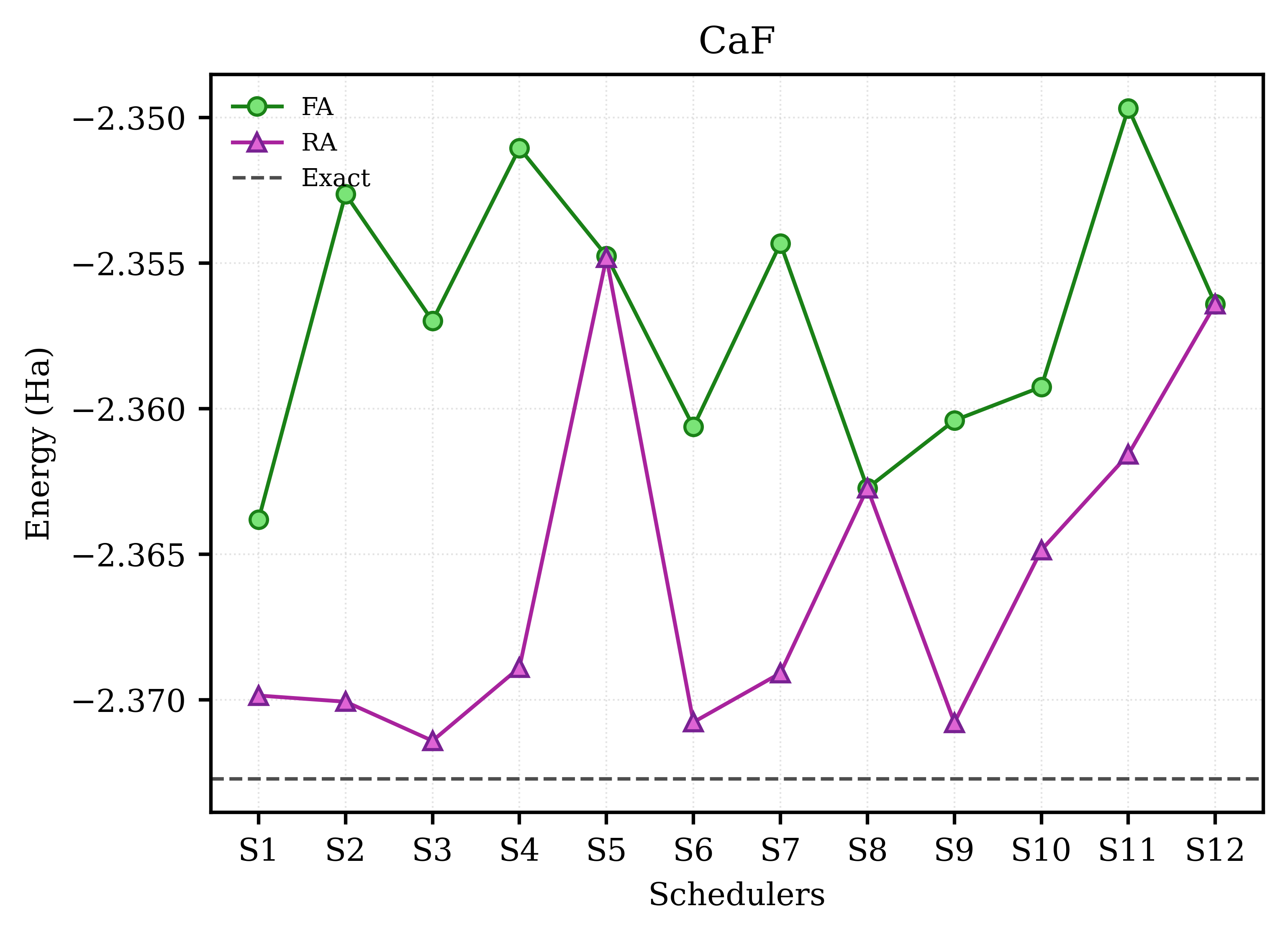}\\
  \footnotesize (c) CaF $(3e,8o)$
\end{minipage}\\[4pt]
\begin{minipage}{0.31\textwidth}
  \centering
  \includegraphics[width=\linewidth]{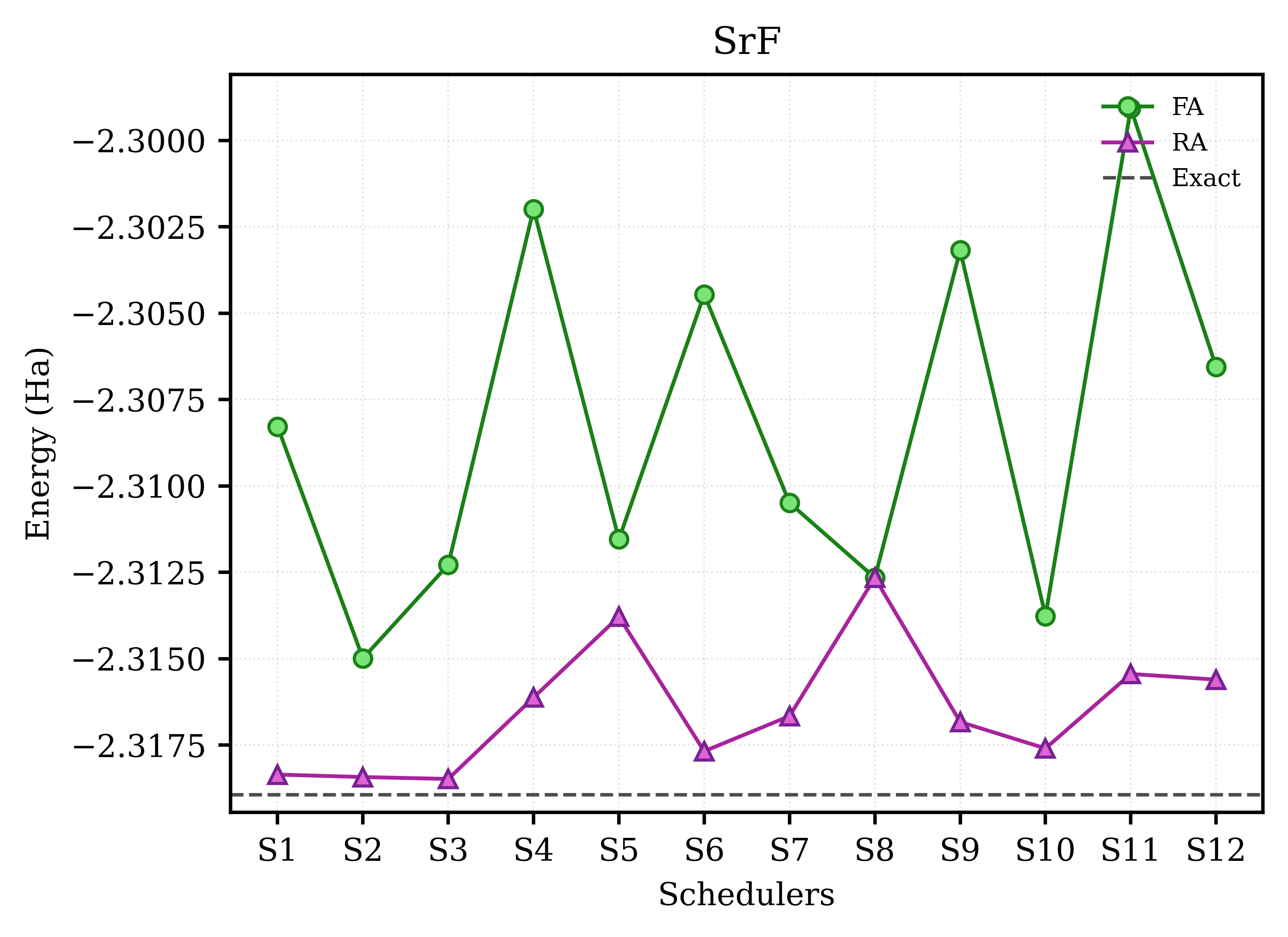}\\
  \footnotesize (d) SrF $(3e,8o)$
\end{minipage}\hfill
\begin{minipage}{0.31\textwidth}
  \centering
  \includegraphics[width=\linewidth]{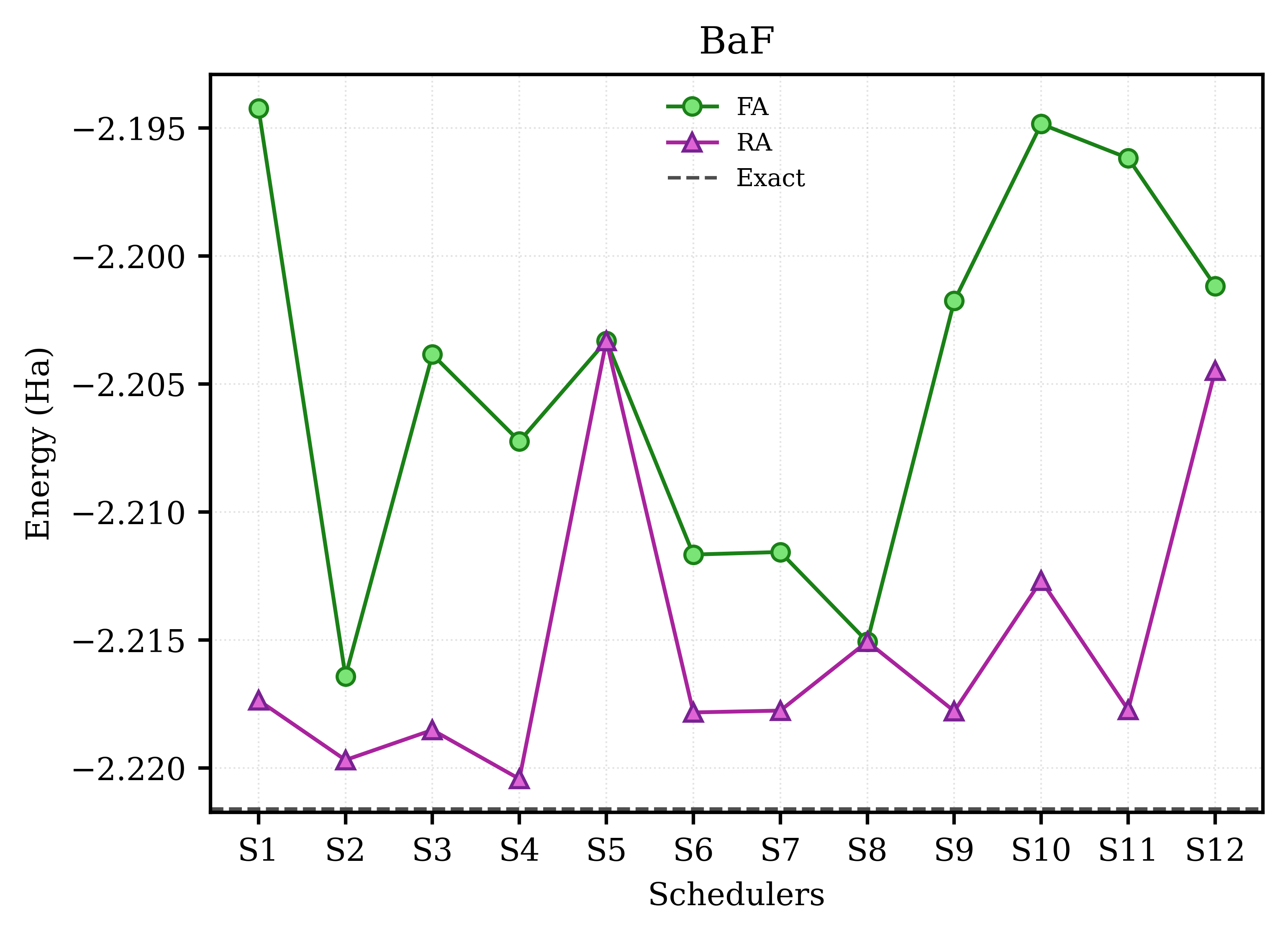}\\
  \footnotesize (e) BaF $(3e,8o)$
\end{minipage}
\caption{Molecule-resolved ground-state energy profiles across schedules $S_1$--$S_{12}$ for the $(3e,8o)$ active space. Green circles denote forward annealing (FA), purple triangles denote reverse annealing (RA), and the dashed black horizontal line marks the exact Full Configuration Interaction (FCI) ground-state reference energy.}
\label{fig:supp_energy_3e8o}
\end{figure*}

Figure~\ref{fig:supp_energy_3e8o} displays the variational ground-state energies obtained after applying RQA for each of the twelve schedules in the $(3e,8o)$ active space. The profiles are plotted alongside the forward-annealing baseline and the exact FCI ground-state reference energy for all five alkaline-earth monofluorides (BeF, MgF, CaF, SrF, and BaF).

Across all five molecules, reverse annealing generally lowers the energy relative to the forward-annealing initialization and moves it toward the exact FCI reference. Deep-reversal schedules such as S1, S4, and S11 yield strong improvements, with BeF reaching within $0.305$~mHa of the exact FCI energy.

\subsection{Molecule-Resolved Energy Profiles in the $(7e,14o)$ Active Space}

\begin{figure*}[t]
\centering
\begin{minipage}{0.31\textwidth}
  \centering
  \includegraphics[width=\linewidth]{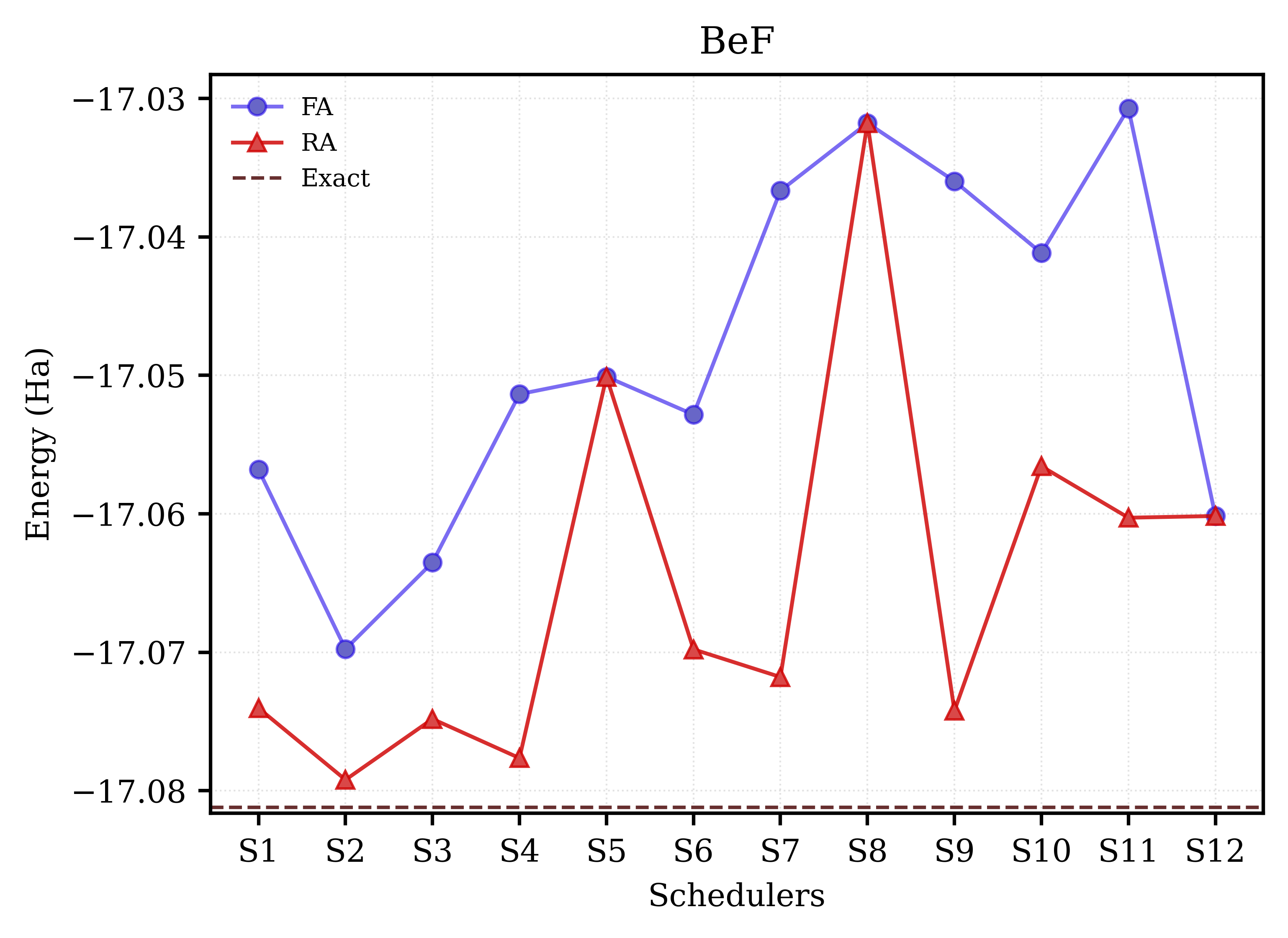}\\
  \footnotesize (a) BeF $(7e,14o)$
\end{minipage}\hfill
\begin{minipage}{0.31\textwidth}
  \centering
  \includegraphics[width=\linewidth]{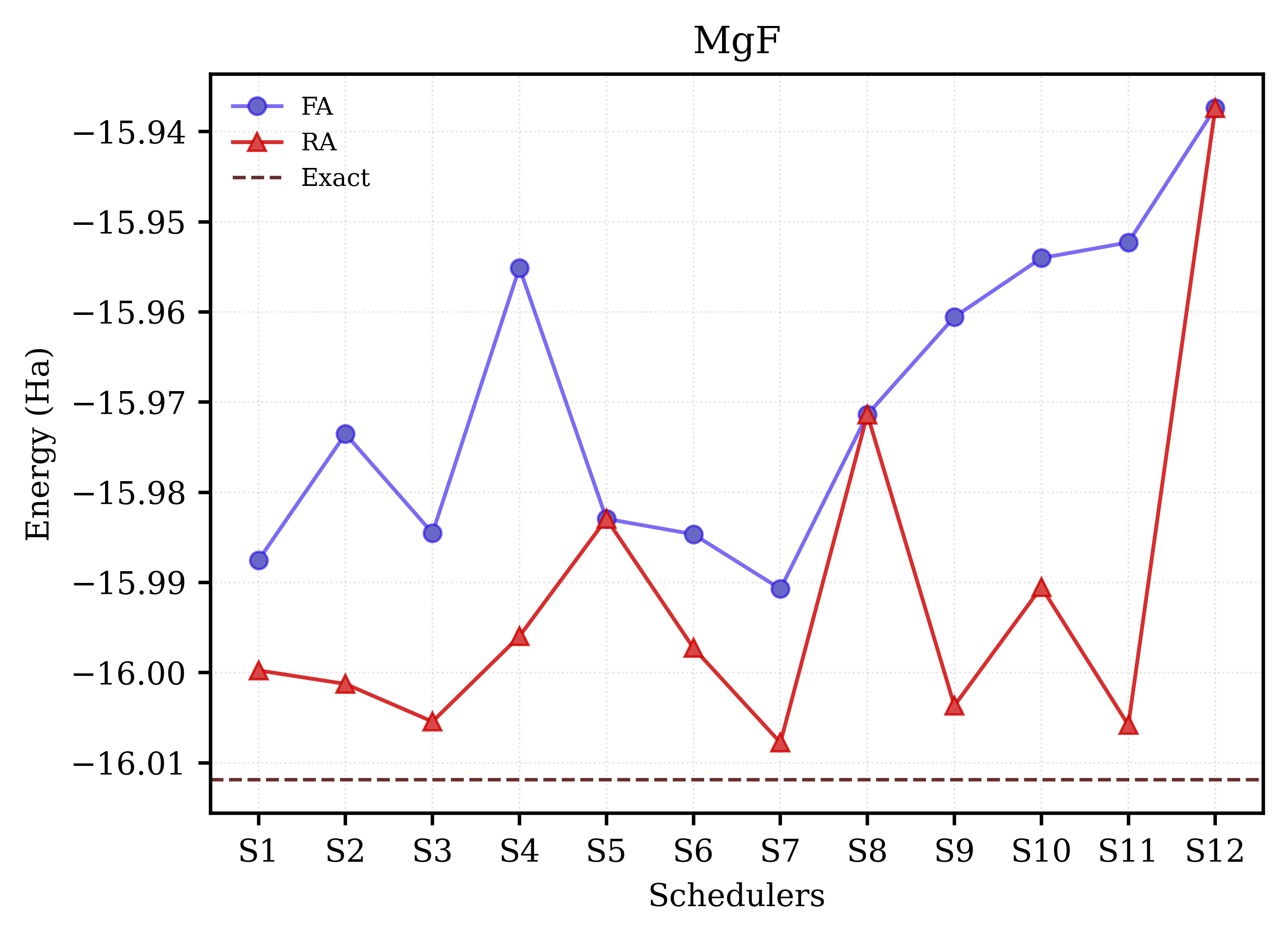}\\
  \footnotesize (b) MgF $(7e,14o)$
\end{minipage}\hfill
\begin{minipage}{0.31\textwidth}
  \centering
  \includegraphics[width=\linewidth]{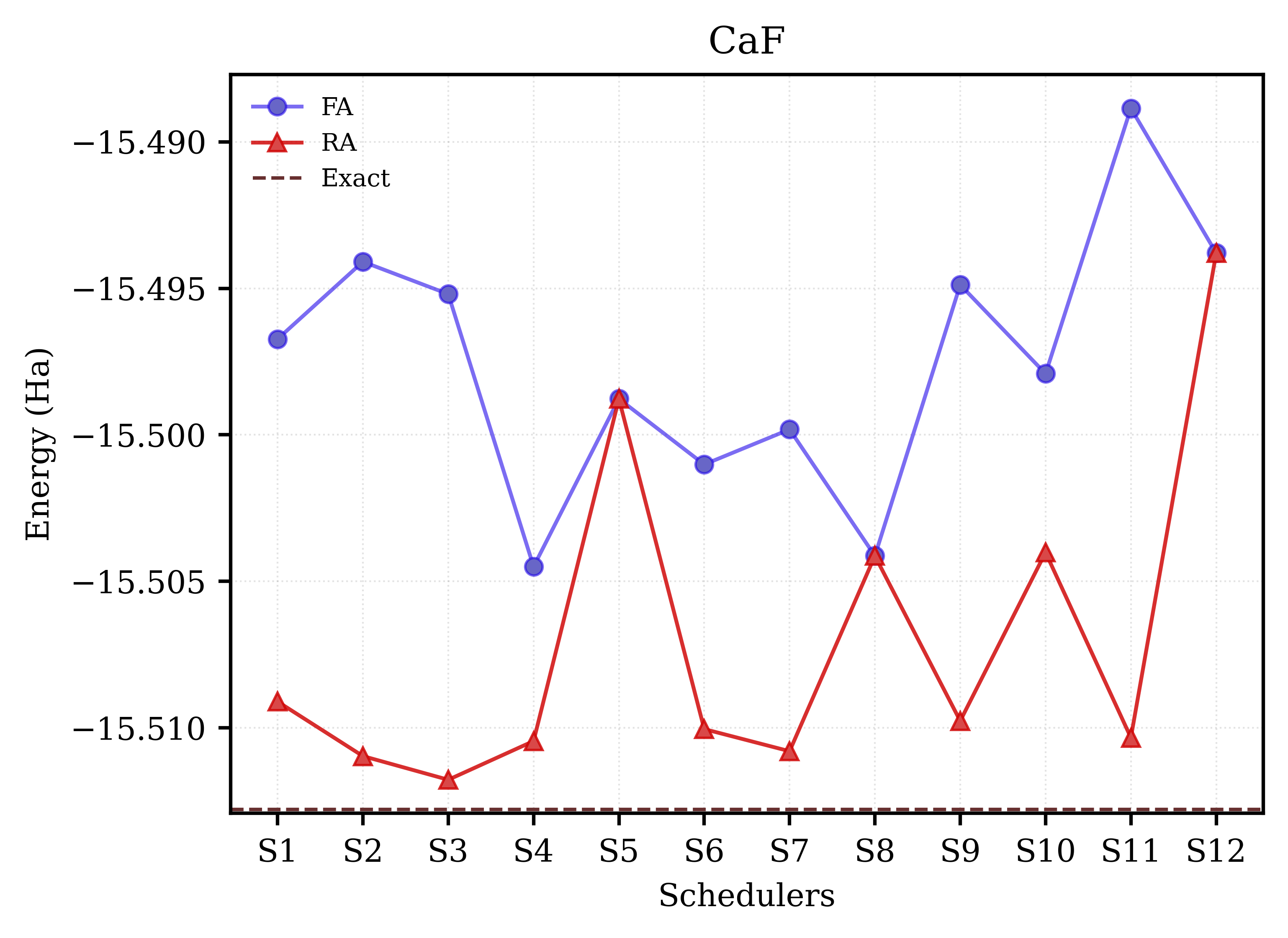}\\
  \footnotesize (c) CaF $(7e,14o)$
\end{minipage}\\[4pt]
\begin{minipage}{0.31\textwidth}
  \centering
  \includegraphics[width=\linewidth]{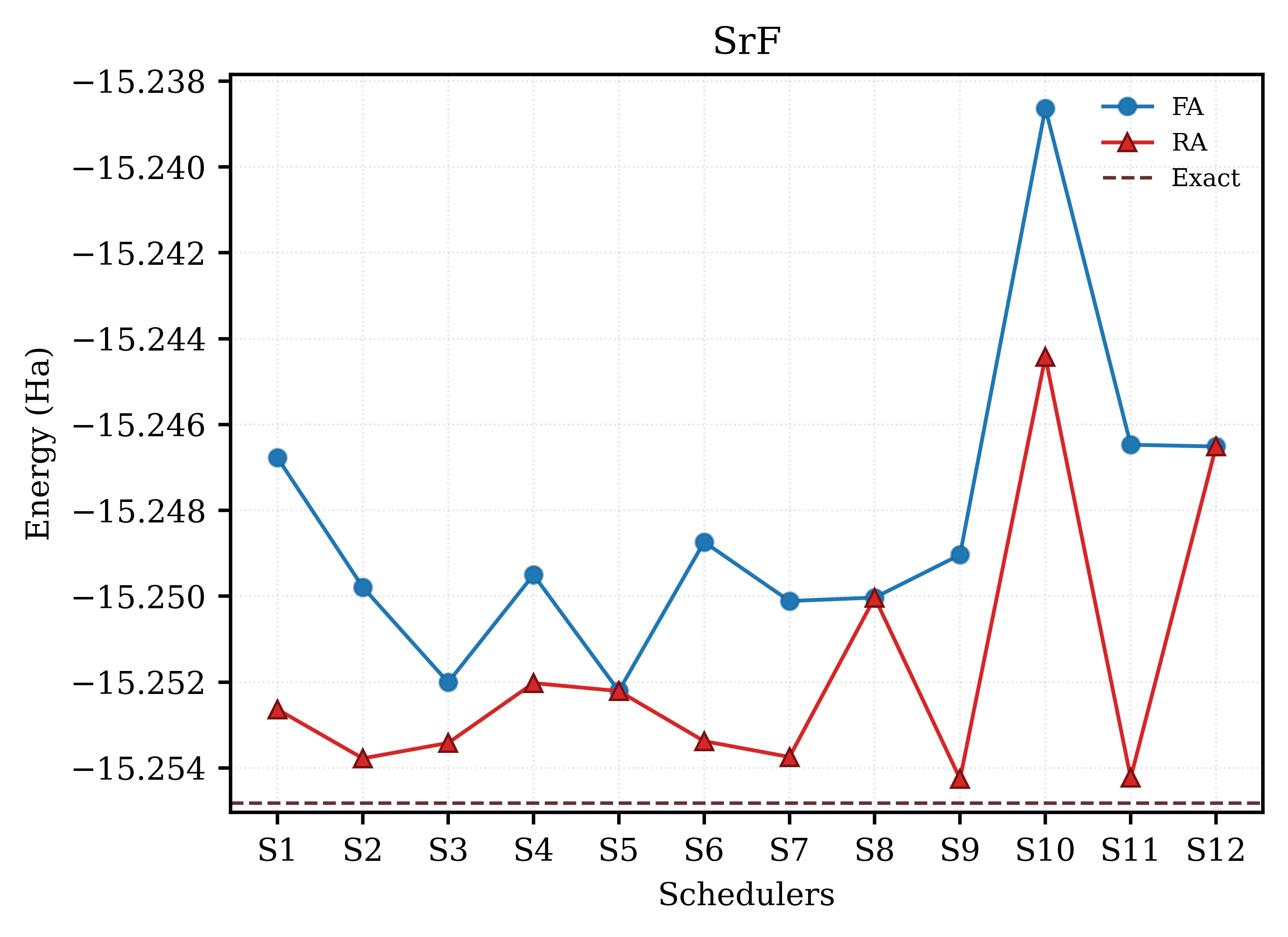}\\
  \footnotesize (d) SrF $(7e,14o)$
\end{minipage}\hfill
\begin{minipage}{0.31\textwidth}
  \centering
  \includegraphics[width=\linewidth]{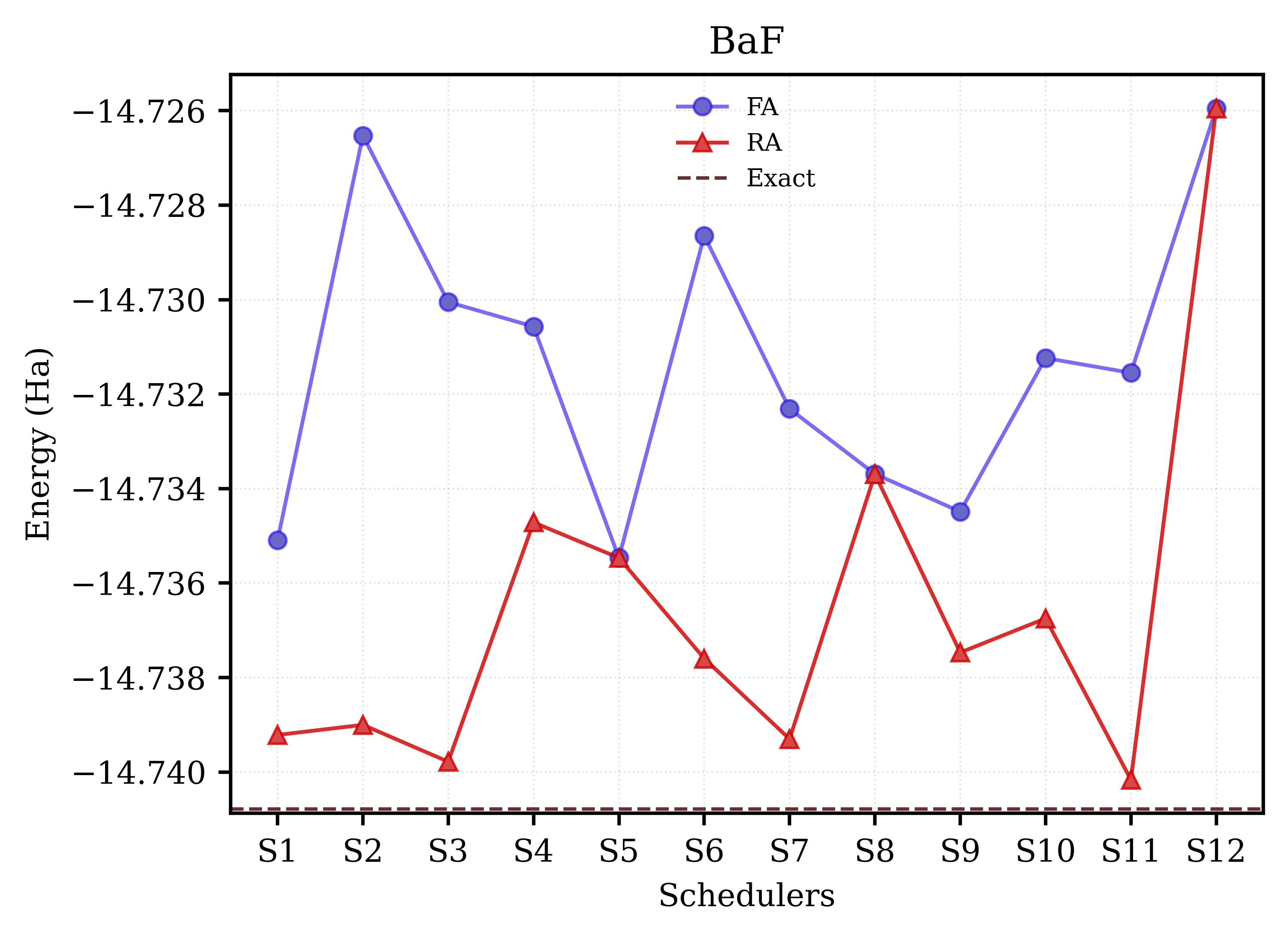}\\
  \footnotesize (e) BaF $(7e,14o)$
\end{minipage}
\caption{Molecule-resolved ground-state energy profiles across schedules $S_1$--$S_{12}$ for the $(7e,14o)$ active space. Blue circles denote forward annealing (FA), red triangles denote reverse annealing (RA), and the dashed black horizontal line marks the exact Full Configuration Interaction (FCI) ground-state reference energy.}
\label{fig:supp_energy_7e14o}
\end{figure*}

Figure~\ref{fig:supp_energy_7e14o} shows the corresponding schedule-resolved energy profiles for the enlarged $(7e,14o)$ active space. Across the sixty molecule--schedule evaluations, reverse annealing matches or lowers the forward-annealing energy in the reported results. SrF reaches within $0.596$~mHa of the exact FCI benchmark, and the mean residual-error reduction is approximately $88\%$ relative to the forward-annealing baseline.

\subsection{Complete Schedule Performance Summary}

Table~\ref{tab:supp_best_schedule} lists the best schedule identified for each molecule in both active spaces, together with the corresponding signed relative energy change $\Delta E$ and absolute residual error relative to exact FCI.

\begin{table}[t]
\centering
\caption{Optimal reverse-annealing schedules for each molecular radical. The percentages denote signed relative energy changes, and $\delta=|E-E_{\rm FCI}|$ is the final residual error.}
\label{tab:supp_best_schedule}
\begin{ruledtabular}
\footnotesize
\setlength{\tabcolsep}{3pt}
\begin{tabular}{lcc}
Molecule & Schedule ($3e,8o$) & $\delta$ (mHa) \\
\colrule
BeF & S1 ($-0.554\%$)  & 0.305 \\
MgF & S11 ($-1.292\%$) & 5.596 \\
CaF & S4 ($-0.759\%$)  & 3.825 \\
SrF & S11 ($-0.712\%$) & 3.492 \\
BaF & S1 ($-1.054\%$)  & 4.246 \\
\colrule
Molecule & Schedule ($7e,14o$) & $\delta$ (mHa) \\
\colrule
BeF & S9 ($-0.225\%$)  & 6.966 \\
MgF & S11 ($-0.335\%$) & 6.065 \\
CaF & S11 ($-0.139\%$) & 2.447 \\
SrF & S11 ($-0.051\%$) & 0.596 \\
BaF & S2 ($-0.085\%$)  & 1.778 \\
\end{tabular}
\end{ruledtabular}
\end{table}

\section{Fixed-Point Encoding Scheme}
\label{app:fixed_point_encoding}

Each variational coefficient $c_i \in [-1, 1]$ is represented using $K=10$ binary variables via $c_i = \sum_{\alpha=1}^{K-1} 2^{\alpha-K} q_\alpha^i - q_K^i$, providing a discretization resolution of $\Delta c = 2^{-9} \approx 1.95 \times 10^{-3}$. This choice balances variational precision against total QUBO dimension ($N \times K$) and physical chain lengths under minor embedding. The identical encoding was used across all five molecules (BeF, MgF, CaF, SrF, and BaF) and both active spaces [$(3e,8o)$ and $(7e,14o)$] in both forward and reverse annealing.

\section{D-Wave Hardware and Reverse-Annealing Protocol}
\label{app:dwave_protocol}

Calculations were executed on the D-Wave Advantage system 4.1 QPU featuring Pegasus graph connectivity. Logical variables were mapped onto physical qubits using minor embedding with majority-vote chain-break resolution. Forward annealing used 1000 reads per problem instance and an annealing time of $20\,\mu\mathrm{s}$. The concatenated forward-annealing configuration served as the classical initial state for reverse annealing.

Twelve schedules (S1--S12) were evaluated, comprising five pause-type schedules (S1--S5, durations $30$--$60\,\mu\mathrm{s}$) and seven V-shaped schedules (S6--S12, duration $20\,\mu\mathrm{s}$), probing reversal depths $s' \in [0.35, 0.65]$ with 1000 reads each. Sampled configurations were decoded and refined via RQA of the continuous QAE functional to restore inter-block couplings truncated during sub-QUBO partitioning. Final energies were benchmarked against forward-annealing baselines and exact Full Configuration Interaction (FCI) diagonalization references.

\end{document}